\documentstyle[floats,aps,aipbook]{revtex}
\begin{document}
% Enter the title in the curly braces below.
\title{Long and Short GRB}
% Enter the author list in the curly braces below.
\author{J. I. Katz and L. M. Canel}
% Enter the address list in the curly braces below.
\address{Department of Physics and McDonnell Center for the Space Sciences\\
Washington University, St. Louis, Mo. 63130}
% Correspondence between authors and address should be established by
%  manually inserting superscript symbols. See aipsamp.tex for an example.
\maketitle
\begin{abstract}
We report evidence from the 3B Catalogue that short ($T_{90} < 10$ s) and
long ($T_{90} > 10$ s) GRB represent different populations and processes:
Their spectral behavior is qualitatively different, with short bursts
harder in the BATSE range, but chiefly long bursts detected at higher
photon energies; $\langle V/V_{max} \rangle = 0.385 \pm 0.019$ for short GRB
but $\langle V/V_{max} \rangle = 0.282 \pm 0.014$ for long GRB, differing by
$0.103 \pm 0.024$.  Long GRB may be the consequence of accretion-induced
collapse, but this mechanism fails for short GRB, for which we suggest
colliding neutron stars.
\end{abstract}

% body of paper here

\section*{Introduction}

The durations of ``classical'' gamma-ray bursts (GRB) are bimodally
distributed and are anti-correlated with their spectral hardness as measured
by BATSE \cite{K93}.  Little attention has been paid to the cause of this
division of GRB into two classes, long and short.  The simplest
interpretation is that they are all of similar origin, but have different
values of one or more parameters, which are bimodally distributed.  Most
models of GRB permit a wide range of parameters.  The observed
anti-correlation of duration with spectral hardness is naturally obtained in
fireball-debris-shock models \cite{RM92,MR93,K94a} in which a higher Lorentz
factor $\gamma$ leads to a shorter GRB with a higher characteristic
synchrotron frequency $\nu_c$.  Bimodality has been attributed \cite{K94a}
to the well-known bimodal distribution of interstellar density.

We re-examine this question.  The most popular fireball models have been
based on coalescing neutron stars \cite{E89}.  However, calculations
\cite{R96,Ma96} indicate that this process is nearly adiabatic and does not
produce sufficient heating or neutrino emission to create an energetic
fireball.

Accretion-induced collapse (AIC) of a bare degenerate dwarf has been
calculated \cite{D92} to produce sufficient neutrino flux to power a
fireball.  AIC produces a $\sim 10$ s neutrino burst, as observed from
SN1987A (the presence of a stellar envelope turns the neutrino energy into a
supernova, while its absence permits a relativistic fireball).  The duration
of neutrino emission is a lower bound on the duration of the resulting GRB
because the subsequent shock interaction, particle acceleration and
radiation can stretch the observed GRB (as will the cosmological redshift),
but cannot so readily shorten it.

AIC can explain only GRB with $T_{90} > 10$ s; shorter GRB require a
different process.  Both classes of events occur at cosmological distances
\cite{K93} and both probably involve neutrino fireball-debris-shock
interactions, explaining their similar phenomenology, but the origin of the
neutrinos must be different.

\section*{3B Hardness and Spectral Data}

If long and short GRB are produced by two distinct physical processes,
members of these two classes may have different spectral properties and
spatial distributions.  This hypothesis can be tested with data in the 3B
catalogue \cite{Me96}.  We find that it is confirmed by qualitative
differences in the spectral properties of long and short GRB and by
quantitative differences in their spatial distributions.

The BATSE hardness ratio is a measure of the spectral slope in the range
50--300 KeV.  Some of the entries in the 3B catalogue were also detected by
COMPTEL, EGRET and OSSE, which indicate the presence of energetic photons
above the BATSE band.  The data are summarized in Table \ref{1}.
\begin{table}
\caption{BATSE hardness and higher energy detections; nominal sensitivity
ranges are indicated.  Data are from 3B Catalogue.}
\label{1}
\begin{tabular}{lccc}
 &$T_{90} < 1$ s&$1\ {\rm s} < T_{90} < 10$ s& $10\ {\rm s} < T_{90}$\\
\tableline
BATSE Hardness Ratio $> 10$ & 18 & 3 & 1 \\
COMPTEL Detections (1--30 MeV) & 2 & 2 & 20 \\
OSSE Detections (0.06--10 MeV) & 0 & 0 & 2 \\
EGRET Detections (20--30,000 MeV) & 0 & 0 & 6 \\
\end{tabular}
\end{table}

Photons of energies $> 1$ MeV are detected almost
exclusively from long ($T_{90} > 10$ s) GRB.  This is opposite to the
behavior expected from an extrapolation of the hardness measured at lower
photon energies by BATSE.  Long and short GRB show fundamentally different
spectral behavior, which cannot be explained by variation of a single
parameter, such as $\nu_c$.  We conclude that they are intrinsically
different objects, involving different physical processes, rather than
different parameter ranges of a single class of event.

If long and short GRB have distinct physical origins they may have distinct
spatial distributions, although this is not required.  Table \ref{2}
presents the results of an analysis of $C/C_{min}$ data in the 3B Catalogue
\cite{Me96}.  The whole Catalogue analysis shows that $\langle V/V_{max}
\rangle$ for long and short GRB differ by $4.3\,\sigma$.  This confirms the
hypothesis that long and short GRB have different spatial distributions and
therefore different physical origins.

\begin{table}
\caption{$\langle V/V_{max} \rangle$ for short and long GRB.  Data from 3B
Catalogue.}
\label{2}
\begin{tabular}{lcccc}
 & Whole Catalogue & 64ms Data & 256ms Data & 1024ms Data \\
\tableline
$T_{90} < 10$ s & $0.385 \pm 0.019$ & $0.383 \pm 0.021$ & $0.373 \pm 0.027$
& $0.391 \pm 0.022$ \\
$T_{90} > 10$ s & $0.282 \pm 0.014$ & $0.370 \pm 0.018$ & $0.305 \pm 0.017$
& $0.276 \pm 0.014$ \\
Difference & $0.103 \pm 0.024$ & $0.013 \pm 0.028$ & $0.069 \pm 0.032$
& $0.114 \pm 0.026$ \\
\end{tabular}
\end{table}

Subdivision of the data on the basis of integration times shows that this
effect is produced by smoothly rising GRB, which do not trigger the
detector when short integration times (64 ms) are used.  We can predict that
if these ``smooth risers'' could be separated from other long GRB they would
have even smaller $\langle V/V_{max} \rangle$, and would be a pure AIC
population, uncontaminated by GRB which have an intrinsically short time
scale (and hence rapid rise) but which are stretched to long $T_{90}$
because they radiate slowly.

\section*{Long GRB}

Any model of long GRB must explain their $> 1$ MeV emission by a process
distinct from that which produces their softer emission.  Such a model was
developed \cite{K94b} for the extraordinarily intense 3B940217, which was
also very long ($T_{90} = 150$ s) as measured by BATSE, and produced
photons of energies as high as 18 GeV an hour after the initial burst
\cite{H94}, but had a BATSE hardness ratio of 3.83, roughly average.  The
energetic gamma-rays were attributed to $\pi^0$ decay or Compton scattering
by energetic electrons and positrons (themselves produced by $\pi^\pm$ decay)
resulting from relativistic nuclei (fireball debris) entering a dense cloud
of circumfireball matter.  We now suggest that such a model is applicable to
all long GRB, though the density and geometry of the cloud will necessarily
vary from event to event, as will therefore the efficiency of production of
energetic gamma-rays.

The cloud was attributed \cite{K94b} to excretion by the progenitor of one of
the neutron stars.  That specific scenario can be replaced by one of AIC;
when matter flows into an accretion disc surrounding the degenerate dwarf a
fraction $f$ of it is excreted from the disc and the binary.  This is
inevitable; matter accreting onto the dwarf must give up nearly all its
angular momentum, which flows viscously outward in the accretion disc.
Conservation of angular momentum gives $f = 1 - ({r_{RCR} / r_{LSO}})^{1/2}$
where $r_{RCR}$ is the Roche circularization radius \cite{K73} and
$r_{LSO}$ is the radius of the last stable disc orbit \cite{B74}, from
which mass peels off the disc and is lost; $f \approx 0.5$, almost
independent of the binary mass ratio.

A degenerate dwarf cannot accrete hydrogen-rich matter faster than $3 \times
10^{-7} M_\odot$/yr because the Eddington limit applies to its thermonuclear
luminosity.  As a result, AIC is likely to be preceded by a period of
accretion of at least $10^7$ yr.  The circumfireball cloud must be rather
small ($< 10^{15}$ cm) \cite{K94b} in order that it
be dense enough for collisional interaction with the relativistic debris;
for energetic collisional gamma-rays detected simultaneously with a 30 s
GRB the time of flight suggests a size $\sim 10^{12}$ cm, but this may be an
underestimate (by a factor up to $\gamma^2$) if the relativistic particles
are moving radially outward at the time of collision.

Even the largest possible cloud is much too small to be freely expanding
over an accretion time of $10^7$ yr.  It could be gravitationally bound in
orbits outside the binary orbit.  Alternatively, accretion of helium or
carbon-oxygen matter could proceed much faster because of the
reduced thermonuclear energy release, efficient neutrino cooling (in burning
of carbon and heavier elements) and the difficulty of igniting these fuels.
Accretion of heavier elements resembles degenerate dwarf coalescence more
than conventional mass transfer, and might be rapid enough (gravitational
radiation-driven coalescence lasts $\sim 30$ yr) that escaping matter would
still be close and dense when the final collapse occurred.

Apart from their gamma-ray emission, such events might resemble supernovae
as energy deposited in the escaping matter is thermalized and radiated.
The predicted gravitational wave emission of a long GRB, produced by AIC, is
$\sim 10^{-9} M_\odot c^2$ \cite{K80,BH96}, or even less if no
matter is expelled.

\section*{Short GRB}

Short GRB require a new mechanism.  The requirement of producing $10^{51}$
erg of soft gamma-rays points to a catastrophic event involving one or more
neutron stars; half the energy must be released in 10 ms in at least a few
GRB.  We suggest the collision of two neutron stars, probably occurring in a
very dense cluster of stars.  Such processes were suggested \cite{G65} as
the origin of quasars.  It is now considered likely that many galaxies
possess massive black holes at their centers, which plausibly grew from
dense clusters of stars.  When the density becomes high collisions become
frequent, and lower density stars are disrupted, leaving only neutron stars
and black holes.

Colliding neutron stars have kinetic energy at impact $\approx$ 150--180
MeV/nucleon and speed (with respect to their center of mass) of $\approx$
0.52--0.55$c$.  This is mildly supersonic, and the shock-heated matter with
$k_B T \sim 100$ MeV is a copious source of neutrinos.  Expelled matter
cools by adiabatic expansion on the time scale $r/v \sim 0.1$ ms, permitting
very short GRB.  Matter retained by the neutron stars may radiate on the
time scale of neutrino cooling, perhaps terminated by collapse to a black
hole.

A cluster of radius $R$, containing $N$ neutron stars each with mass $M$ and
radius $r$, has an evaporation time
$$t_{ev} \approx {200 N \over \ln N} t_{cr}, \eqno(1)$$
where $t_{cr} \equiv (R^3/GMN)^{1/2}$ is the crossing time.  The collision
time is
$$t_{coll} \approx {R \over r} t_{cr}, \eqno(2)$$
where the cross-section, allowing for gravitational focusing, is $\approx
rR/N$.  The total collision rate is
$$\nu_{coll} \sim \left({GM r^2 N^3 \over R^5}\right)^{1/2} \sim 10^{19}
{N^{3/2} \over R^{5/2}} {\rm s}^{-1}. \eqno(3)$$

A hypothetical cluster with $N = 10^8$ and $R = 10^{18}$ cm (virial velocity
$\approx 2 \times 10^8$ cm/s) has a collision rate $\sim 10^{-14}$
s$^{-1}$ and a lifetime of $\sim 10^{19}$ s.  About $10^9$ such clusters
would be required to produce the observed $10^{-5}$ short GRB s$^{-1}$
within $z \sim 1$; this number is comparable to the $\sim 10^9$ galaxies in
that volume, and we cannot exclude that such clusters are commonly found at
the centers of galaxies.

A much smaller number of more active clusters may be the sources of short
GRB, in which case repetitions may be found.  The present upper bound
\cite{Me95} on the repetition rate of GRB is not stringent.

A simple estimate shows that the gravitational radiation emitted in the
collision of two neutron stars is $\sim 10^{-2} GM^2/r \sim 10^{51}$
erg into a broad band around $\sim 3$ KHz.  The wave-train would be very
different from that of coalescing neutron stars.

This research has made use of data obtained through the Compton Gamma-Ray
Observatory Science Support Center Online Service, provided by the
NASA-Goddard Space Flight Center.  We thank R. Kippen and C. Kouveliotou for
discussions and NASA NAGW-2918 and NAG-52862 and NSF AST 94-16904 for support.

\end{document}